# 10 Inventions on scrolling and scrollbars in Graphical User Interface
## A TRIZ based analysis

**Umakant Mishra**

Bangalore, India

http://umakantm.blogspot.in

**Contents**

## 1. Introduction

The processing and memory capability of a computer allows very large documents, tables and images to be created and manipulated using various software programs. However, the size of the display screen is typically very limited and not enough to display these large documents. When the data or document being displayed within the window is larger than the display area or window, some data remains hidden beyond the boundary of the window. There is a need for a mechanism to display and edit large documents in the limited screen space or window.

- One option to display large documents within the screen is to increase the screen size. But large monitors or display devices are very expensive. Besides there is a physical limitation to expand the monitor size beyond a limit.

- Another solution to this problem is to reduce the size of the picture or document. But reducing the size of the picture reduces clarity of viewing.

- The most popular solution to this problem is scrolling. Scrolling refers to selectively moving this hidden portion of data inside the display area.

## 1.1 Situations needing scrolling operation

- Sometimes the data or document is too large to be shown within the display area in its entirety. For example, a large graphic image can only be shown partially within the display area. This situation needs a scrolling mechanism to view different parts of the document.

- Scrolling may be used for container windows to display a large number of objects, which do not fit within the size of the window.

- Sometimes the user may like to zoom in or magnify an object to get a larger view. The magnified object may not fit within the size of the window, which needs scrolling.

## 1.2 Different methods used for scrolling

- The scrolling mechanism may include a vertical scroll bar or a horizontal scroll bar or both to move the contents of the documents up and down or left and right.

- The navigation keys on the keyboard may be used to move up, down, left and right.

- There may be navigation buttons on the screen representing the navigation keys on the keyboard. The user can click these buttons to scroll the screen.

- In some cases just moving the mouse to any direction may implement a scrolling of the screen to that that direction.

- Another method is to drag the screen by using the "thumb", as popularly used with a pdf document. This action is analogous to moving a flat piece of paper around on a desk with one's hand.

- Scrolling may be achieved through eyeball tracking in a hands-free environment where the user does not have hands or wants to use hands for other activities.

## 2. Inventions on scrolling and scrollbars

Although a conventional scrollbar does its function well in most cases, there are situations, which demand improvements in the mechanism. For example, where the size of the window is too small, the presence of a scrollbar on the screen reduces the size further. Reducing the width of the scrollbar may solve this problem to some extent but that would require higher precision of cursor movement and cause more errors in scrolling. There are various inventions to overcome these problems. This article will illustrate ten interesting inventions on scrolling selected from US patent database.

### 2.1 Auto-scrolling during a drag and drop operation (5611060)

**Background problem**

During a drag and drop operation, the destination may not be visible in some instances because of the limited size of the display window. In that case it is difficult to drag the object precisely on to the destination place. In a conventional mechanism the user can place the mouse indicator on the scrollbar to scroll up or down to locate the destination. But this method of scrolling while dragging often causes undesirable scrolling and lead to user frustration.

**Solution provided by the invention**

Patent 5611060 (invented by Belfiore et al., assigned by Microsoft Corporation, issued March 1997) provides a scrolling during drag and drop operation where the scroll is determined by the location and speed of the mouse indicator. When the mouse indicator is over a predefined area of the window (say at the border) the system compares the speed of the mouse indicator to a predetermined threshold and scrolls the window if the speed is less than the predetermined threshold.

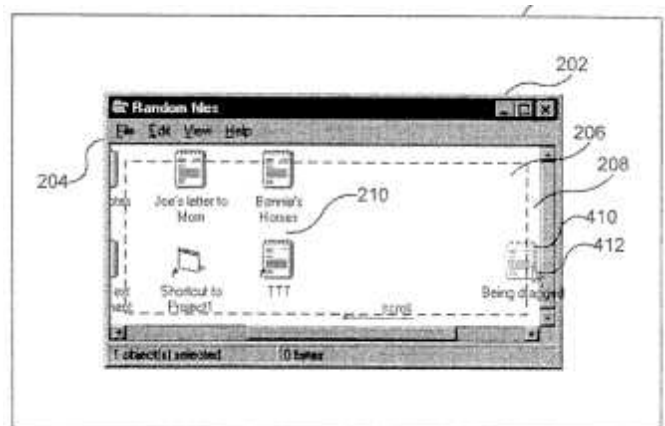

The invention includes a calculation component, a location component and a scrolling component. The calculation component calculates the speed of the mouse pointer while dragging the object. The location component determines whether the screen object is located over a predefined area of the window.

## TRIZ based analysis

The invention provides scrolling of the target window during drag and drop operation (Principle-15: Dynamize).
The speed and direction of scrolling of the window depends on the location component and calculation component. (Principle-40: Composite).

## 2.2 Differently magnified interlocked windows with automatic scrolling (5625782)

### Background problem

In some graphic editing tools, there are two windows, one window shows the characters/figures in a small size for preview and the other window shows the characters/figures in large size for editing. This method of simultaneous viewing and editing of the characters/figures improves the efficiency of the operator.

But this method has a drawback. As the window1 showing the miniature view occupies some screen space, the window2 gets less screen space, which requires frequent scrolling by the user. To solve the above problem the Japanese patent No 270384 (1992) proposes to display a rectangular body on window1. By moving the rectangular body on window1, the contents of window2 are scrolled.

### Solution provided by the invention

Patent 5625782 (invented by Soutome et al., assigned by Hitachi Ltd, issued Apr 1997) discloses a solution to the above problem. The invention displays a first window and a second window in a non-overlapping manner. The first window contains the reduced size and the second window contents the actual or enlarged size. The input or editing can be carried on from either the first window or the second window. Any editing or cursor movement in the first window reflects corresponding changes in the second window and vice versa.

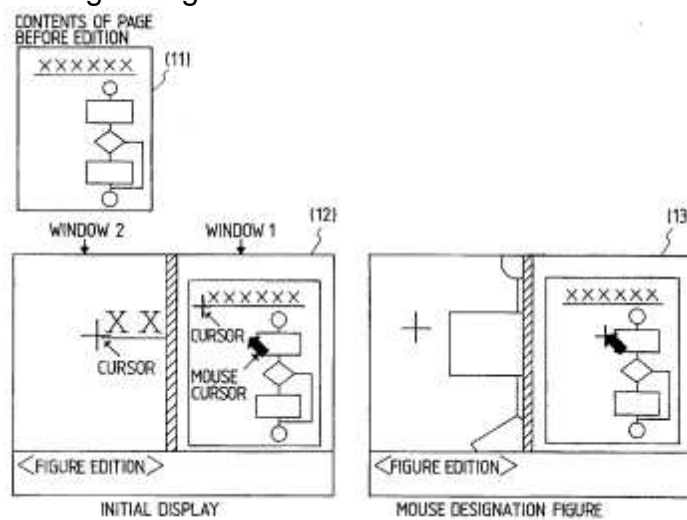

When position of the cursor moved in window1 is not visible in window2, the display position in window2 is automatically scrolled to display the position.

**TRIZ based analysis**

The invention uses a pair of interlocked windows, the editing and cursor movement in any of the window reflects in the other (Principle-26: Copy).
As scrolling is a difficult task, the reduced window is used to move the cursor to the desired place which results in automatic scrolling of the enlarged window and moving the cursor to the desired place (Principle-25: Copy, Principle-15: Dynamize).

**2.3 Method and apparatus for improved scrolling functionality in a graphical user interface utilizing a software tether (5874957)**

**Background problem**

A scrolling function is necessary as most documents or graphical or image files cannot be displayed within a single workspace or within the limited area of the screen. Typically vertical and horizontal scrollbars are used to accomplish the scroll function. But this conventional method of scrolling confuses the operator about his/her location in software object while scrolling. There is a need for an improved scrolling mechanism.

**Solution provided by the invention**

Patent 5874957 (invented by Cline et al., assigned by International Business Machines Corporation, issued Feb 1999) provides an improved scrolling functionality by using a software tether. The tether is a thin line, which does not provide any substantial impediment to the visibility of the software objects displayed in the area.

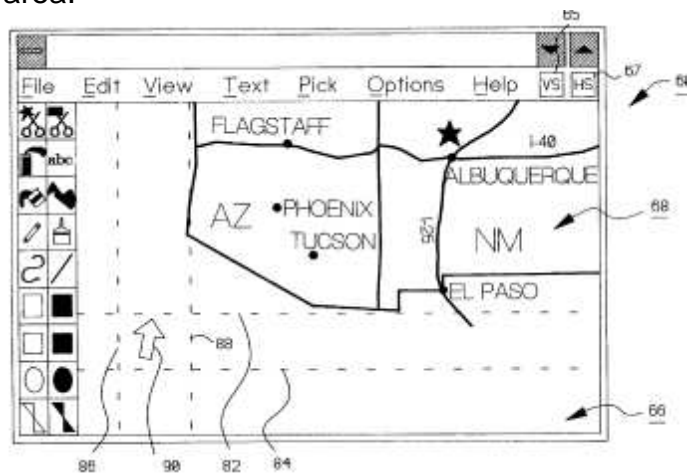

The invention provides a scroll bar activation function which associates scrollbar functionality to the mouse pointer (or similar pointing device). When the scrollbar function is activated the screen scrolls with the movement of the mouse pointer.

**TRIZ based analysis**

The invention provides a tether by which the operator keeps track of his position in the graphics or image being scrolled (Principle-8: Counterweight).
Unlike the conventional scrollbars the tether is a thin line which does not hide any important area of the screen (Principle-30: Thin and flexible).

## 2.4 Method and apparatus for providing feedback while scrolling (6061063)

### Background problem

There are several ways to scroll the contents of the display region. For example, the user may scroll by moving the mouse pointer, or using the scroll bars etc. The user may also drag the screen sometimes called "thumb". The user may use a scroll box to scroll the display screen. But each of them has its limitations. There is a need for a simpler and efficient scrolling mechanism that consistently provides instructional feedback.

### Solution provided by the invention

Patent 6061063 (invented by Wagner et al., assigned by Sun Microsystems Inc, issued May 2000) discloses an invention of efficient scrolling that provides the user with visual and operational clues. The display area may contain a list, a text box, a popup menu or any kind of data. The user is provided with operational clues such as what actions move the list and what actions do not move the list. The invention provides various options to provide operational clues.

(i) In one option a blank space may be used as a visual cue. When the user is at the top of the list or at the bottom of the list the blank space is displayed to indicate that the user cannot scroll further. (ii) In another option the control buttons may be used to communicate the status. For example to control buttons are disabled when the list is scrolled to the top or to the bottom. (iii) Another option is to use the partially visible fields to inform the user that there are additional fields in the partially visible direction.

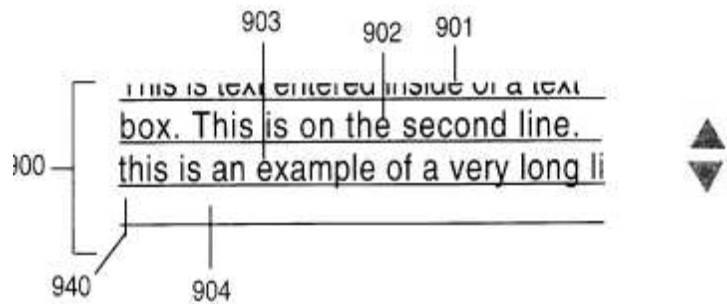

All these methods are applicable to lists, pop-up menu, textbox and other such objects scrolled. (This patent has a continuation patent 6300967, by the same inventor and assignee).

### TRIZ based analysis

The invention intends to help user in scrolling operation by providing informational feedback to the user (Principle-8: Counterweight).

One method is to use a blank field to inform the end of the list. Another method is to change color and disable the scrolling buttons (Principle-32: Color change).
One implementation uses a partially visible field to indicate that there are additional fields beyond the partially visible field (Principle-32: Color change).

## 2.5 Method and apparatus for improved scrolling functionality in a graphical user interface utilizing a transparent scroll bar icon (6069626)

### Background problem

In most application environments the workspaces are larger than the size of the window. In those cases the scroll bars are used to scroll the large workspace vertically and/or horizontally to facilitate viewing the large workspaces. However, the scrollbars occupy some valuable screen space, which can be used for other purposes. How to implement scrollbars without occupying any screen space?

### Solution provided by the invention

Cline et al. disclosed this method of using a transparent scrollbar (Patent 6069626, assigned to IBM, May 2000). The transparent scrollbar can be placed even on the workspace. The transparent scrollbar is preferably smaller in size and movable on the GUI so that it does not create any impediment in viewing the workspace.

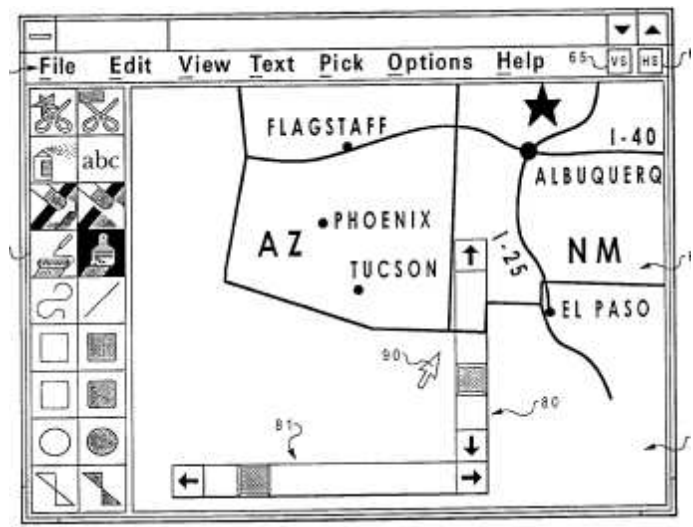

### TRIZ based analysis

The scrollbar should be visible to see the buttons and current screen positions, but it should not occupy any screen space on the GUI. (Contradiction)

The invention provides transparent scrollbars which are placed on the workspace but they don't hinder the visibility of the workspace (Principle-32: Color change).

The transparent scrollbars is preferably smaller in size and capable of moving about the graphical user interface using a pointing device like mouse (Principle-30: Thin and flexible).

## 2.6 Graphical user interface inline scroll control (6181316)

**Background problem**

Currently the window based user interfaces use scrollbars to view large workspaces by moving up and down in a small window. Alhough a scrollbar is a very useful control on the GUI, it consumes fixed amount of space and remains on screen through out the existence of the window. The other disadvantage of a scrollbar is that it requires long pointer movements as it is placed at the side and bottom of the window. There is a need to save the screen space typically occupied by scrollbars and reduce the pointer movements to operate the scrollbars.

**Solution provided by the invention**

Little et al. invented an inline scroll control (Patent 6181316, assigned to IBM, Jan 2001) which places the scroll control directly on the worksheet. The scrolling is achieved by clicking on the up and down indicators. This inline scrolling is designed to reduce the amount of space required on a display device to convey information to the user.

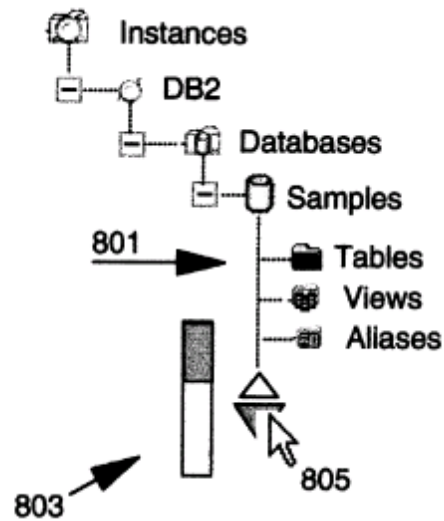

The scrollbars are small in size (like buttons), which requires less pointer movements. Reducing pointer movement is very useful typically in portable computers having less efficient pointing devices like trackballs.

**TRIZ based analysis**

The invention displays the scrollbar on the worksheet itself instead of conventional positioning at the side and bottom of the window (Principle-17: Another Dimension).

The size of the scrollbar is reduced to reduce the pointer movements (Principle-35: Change parameter).

## 2.7 Portable information terminal and information scrolling method for use therewith (6201524)

### Background problem

With the advancement of telecommunication systems, there is a growth in use of portable telephones and pagers for character based communication. As the screen of a pager is very small and there is limited number of control keys, the information scroll is generally permitted in the downward direction only. However, users feel constrained when they can scroll information in a single direction alone. The information should preferably be scrolled up and down to provide convenience.

### Solution provided by the invention

Patent 6201524 (invented by Aizawa, assigned by Sony Corporation, issued Mar 2001) solves the above problem by providing two scroll modes, viz., a line scroll mode and a screen scroll mode, without providing additional control keys.

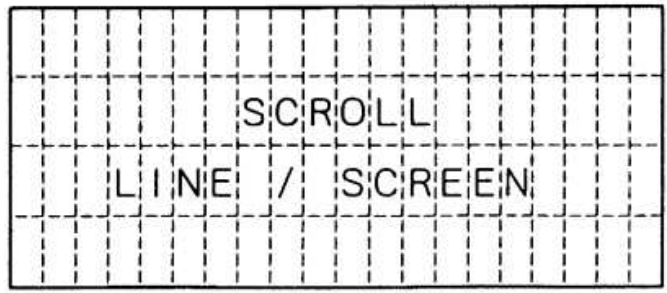

### TRIZ based analysis

The invention provides two types of scrolling, viz., line scrolling and screen scrolling instead of conventional method of providing only one directional scrolling (Principle-17: Another dimension).

## 2.8 Intelligent scrolling (6331863)

### Background problem

Normally in a drag and drop operation the user drags a source object and drops in a destination object/ window. But in such operation the destination object/ window should be visible at the time of operation. When the data displayed within the window is larger than the display area, some part of the data is "hidden". If a selected item is being dragged to a folder that is "hidden" then the prior art methods do not work, as the destination folder is not visible to be dropped.

There are some prior art solutions to this problem. For example, one may drop the object on a temporary place like desktop, scroll the destination folder to make visible and then drag again from the temporary place to drop on the destination folder. Alternatively one may open the destination folder in a second window and drag the object from the source window to the destination window.

All the above methods of solution need the object to be deposited outside the window. It is desirable to provide for scrolling in a window when items have been selected and are being "dragged" to a folder within the same window.

**Solution provided by the invention**

Patent 6331863 (invented by Meier et al., assignee Apple Computer Inc., issued Dec 2001) discloses a context sensitive scrolling or intelligent scrolling. The invention allows selecting one or more items in a window and move the items to a "hidden" destination in the same window. According to the invention, when the user move the selected items to the edge of the window and pauses for a predefined period of time, the window starts scrolling towards that direction thereby making the "hidden" area visible.

(Picture)

The present invention also works for two windows. When the object is dragged from a first source window to a second destination window, placing the cursor at the edge of the second window can scroll the second window.

**TRIZ based analysis**

The invention scrolls the window based on the position of the dragging cursor thereby making the hidden destination visible (Principle-15: Dynamize).

**2.9 Relevance-enhanced scrolling (6339437)**

**Background problem**

When a user gives a search to the World Wide Web search engines the search engine responds with a document containing the search results. In many cases the search engine marks all the occurrences of terms from the query in the document with an HTML </strong> tag. While viewing this result document, the user has to scroll and stop and regular intervals to view the relevance of the content. This problem occurs in browsers, in information retrieval systems and in word processing programs.

**Solution provided by the invention**

Patent 6339437 (invented by Nielsen, assigned by Sun Microsystems Inc, Issued Jan 2002) discloses a method of scrolling based on the presence of a relevant information at different locations in a document.
According to the invention, the searched document can be marked with relevance markers by the search engine. If the search engine is not so equipped the markers can be added locally by the browser.
The relevance markers are typically marked with a <relevance> or <strong> tag. When the user scrolls the document the mechanism looks for the occurrence of the next relevance marker and stops scrolling at the each such occurrence.

**TRIZ based analysis**

The invention uses a relevance marker to scroll to relevant information. The relevant marker is created by marking a <relevant> tag or HTML <strong> tag or similar (Principle-4: Asymmetry).

If the search engine has not added the relevance markers the invention adds relevance markers to the document at the browser level (Principle-3: Local quality).

When the user scrolls the document the scroll operation pauses at every relevance marker and emphasizes the visibility of the relevant information by using color or sound etc. (Principle-21: Skipping, Principle-35: Parameter change).

## 2.10 System and methods for controlling automatic scrolling of information on a display or screen (6351273)

**Background problem**

The standard GUI interface for scrolling is controlled by a pointing device like mouse. But how to scroll the information on the screen in a hands-free environment where the user does not have hands to operate a mouse or wants to use the hands for other activities like using a keyboard. All the previous inventions in this area have various limitations. There is a need for a hands free eye controlled scrolling device for computer systems. There is also a need to provide an automatic scroll control device for automatically scrolling the display of information, text, data, images etc. on a computer display screen to provide a hands-free environment.

**Solution provided by the invention**

Patent 6351273 (invented by Lemelson et al., issued Feb 2002) discloses a method of automatic scrolling of information on a computer display by tracking the position of user's head and eye using a computer-gimbaled sensor.

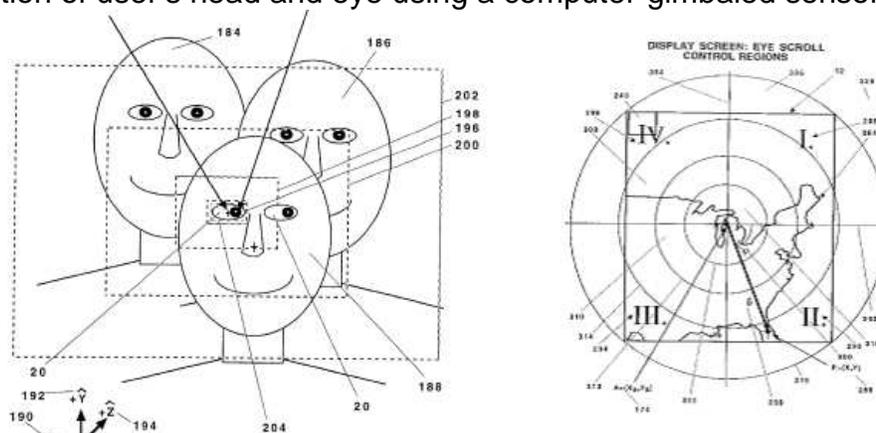

The gimbaled sensor system tracks the eye of the user and an eye gaze direction determining system determines the eye gaze direction. An automatic scrolling system scrolls the screen based on calculated screen gaze coordinates of the eye of the user.

**TRIZ based analysis**

The invention uses an eye tracking mechanism to scroll the screen instead of conventional mouse like pointers (Principle-28: Mechanics substitution).

The invention uses a non-attached mechanism, which means, no attachments are mounted on user's head (Principle-2: Taking out).

The method discloses an automatic scroll control mechanism (Principle-25: Self service).

The method segments the screen control region into a number of concentric circles (Principle-1: Segmentation, Principle-14: Curve).

The invention discloses an algorithm of converting the user's eyeball movement to find screen gaze coordinates (Principle-36: Conversion).

## 3. Summary and conclusion

As we saw the inventions on scrolling tries to overcome the limitations of conventional scrollbars. The future inventions will also try to enhance the important features like:

- Graphical scrolling controls using less screen space.
- Positioning of scrolling controls without obscuring valuable information.
- Scrolling to all directions including diagonal instead of conventional horizontal and vertical scrolling.
- Easy mechanism of scrolling reducing errors in scrolling.
- Easy controlling speed of scrolling.
- Automated scrolling based on user requirement.
- Automated scrolling without using hands.